\documentclass[11pt,twoside]{article}
\usepackage{./asp2014}

\aspSuppressVolSlug
\resetcounters

\begin{document}

\title{IVOA Provenance data model: hints from the CTA Provenance prototype}
\author{Michèle Sanguillon$^1$, Mathieu Servillat$^2$, Mireille Louys$^3$, François Bonnarel$^3$, Catherine Boisson$^2$, Johan Brégeon$^1$}
\affil{$^1$Laboratoire Univers et Particules de Montpellier, Université Montpellier 2, CNRS/IN2P3, France; \email{michele.sanguillon@univ-montp2.fr}}
\affil{$^2$LUTH, Observatoire de Paris, PSL Research University, CNRS, Université Paris Diderot, Sorbonne Paris Cité, France; \email{mathieu.servillat@obspm.fr}}
\affil{$^3$Centre de Données astronomiques de Strasbourg, Observatoire de Strasbourg, Université de Strasbourg, CNRS, France}

% This section is for ADS Processing.  There must be one line per author.
%\paperauthor{Sample~Author1}{Author1Email@email.edu}{ORCID_Or_Blank}{Author1 Institution}{Author1 Department}{City}{State/Province}{Postal Code}{Country}
\paperauthor{Michele Sanguillon}{michele.sanguillon@univ-montp2.fr}{}{LUPM}{}{Montpellier}{}{}{France}
\paperauthor{Mathieu Servillat}{mathieu.servillat@obspm.fr}{orcid.org/0000-0001-5443-4128}{LUTH}{}{Meudon}{}{}{France}
\paperauthor{Mireille Louys}{mireille.louys@unistra.fr}{}{CDS}{}{Strasbourg}{}{}{France}
\paperauthor{François Bonnarel}{francois.bonnarel@astro.unistra.fr}{}{CDS}{}{Strasbourg}{}{}{France}
\paperauthor{Catherine Boisson}{catherine.boisson@obspm.fr}{}{LUTH}{}{Meudon}{}{}{France}
\paperauthor{Johan Brégeon}{johan.bregeon@lupm.in2p3.fr}{}{LUPM}{}{Montpellier}{}{}{France}

\begin{abstract}
We present the last developments on the IVOA Provenance data model, mainly based on the W3C PROV concept. In the context of the Cherenkov astronomy, the data processing stages imply both assumptions and comparison to dedicated simulations. As a consequence, Provenance information is crucial to the end user in order to interpret the high level data products. 
The Cherenkov Telescope Array (CTA), currently in preparation, is thus a perfect test case for the development of an IVOA standard on Provenance information.
We describe general use-cases for the computational Provenance in the CTA production pipeline and explore the proposed W3C notations like PROV-N formats, as well as Provenance access solutions.
\end{abstract}

\section{Introduction}

The International Virtual Observatory Alliance\footnote{\url{http://www.ivoa.net/}, and \url{http://wiki.ivoa.net/twiki/bin/view/IVOA/IvoaDataModel}} (IVOA) has developed several data models to foster interoperability between the various astronomy projects. Those efforts focused mainly on the representation of metadata necessary to describe the content of datasets searched and used by astronomers for their science work.

However, raw data are generally processed, combined, and interpreted in order to lead to a final science-ready data product. Information about the acquisition of raw data and the processing steps in a production pipeline that generates the data products are critical for the user to evaluate data quality, and ensure the correct interpretation of the data products. Such Provenance information is also required to retrace the data processing steps, locate errors and adapt input parameters for a user specific needs.

This is the reason why the IVOA investigates how to model the Provenance of a data set\footnote{\url{http://wiki.ivoa.net/twiki/bin/view/IVOA/ObservationProvenanceDataModel}}, both in terms of observing configuration and of data processing.

\section{The Cherenkov astronomy as a use case}

The Imaging Atmospheric Cherenkov Technique (IACT) is a method to detect very high energy gamma-ray photons in the 50 GeV to 50 TeV range. The IACT works by imaging the very short flash of Cherenkov radiation generated by the cascade of relativistic charged particles produced when a very high-energy gamma ray strikes the atmosphere. This is illustrated in Figure~\ref{fig1}.

\articlefigure{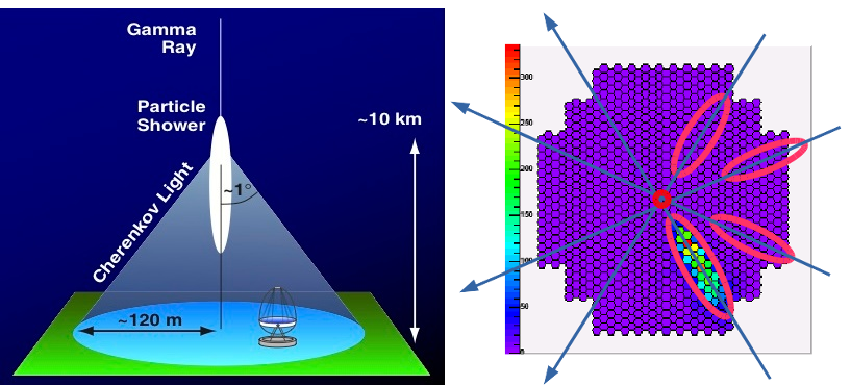}{fig1}{Detection of gamma-rays using Cherenkov telescopes. {\it Left}: schema of a particle shower entering the atmosphere. The Cherenkov light is caught by one or several telescopes. {\it Right}: image of the shower on the camera of a telescope and basic reconstruction of the event direction for four telescopes.}

The Cherenkov Telescope Array (CTA) project is an initiative to build the next generation ground-based very high energy gamma-ray instrument. It will provide a deep insight into the non-thermal high-energy universe \citep[see e.g.][]{CTA2013}. Contrary to previous Cherenkov experiments, it will serve as an open observatory to a wide astrophysics community, with the requirement to propose self-described data products to users that may be unaware of the Cherenkov astronomy specificities.

The CTA is particularly complex to describe because it will be composed of tens of telescopes of three different sizes (4, 12 and 24 meters) that will be equipped with different types of camera. A detailed description of the observing configuration is thus required. Moreover, the processing pipeline includes a reconstruction stage, specific to Cherenkov astronomy, where data from all telescopes are used to model the particle shower and deduce the sky coordinate, time and energy of an incident gamma-ray photon. Photons are thus observed indirectly, and the high level data products are always based on assumptions and comparisons to detailed simulations.

\articlefigure{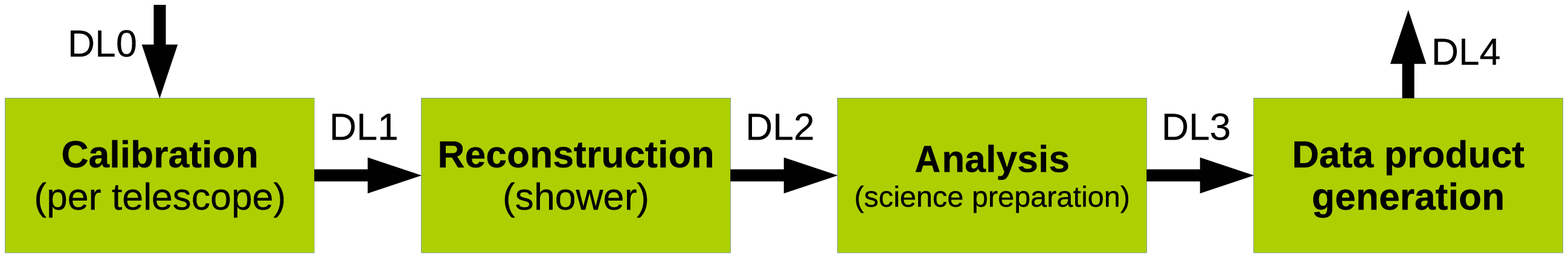}{fig2}{Workflow for the processing activities of Cherenkov astronomical data. Data levels (DL) are indicated between the different pipeline stages}

The stages of the pipeline are shown in Figure~\ref{fig2}, from raw data (DL0) to data products distributed to the community: event lists (DL3), and images, lightcurves and spectra (DL4). DL5 will correspond to catalogs and collections of data products. Data levels are described in more details in e.g. \cite{Contreras2015}. High level data products (DL3 to DL5) will be made available through the Virtual Observatory (VO). To this aim, a data access prototype\footnote{\url{http://voparis-cta-client.obspm.fr/}} has been developed at the Paris Astronomical Data Center to ensure the compatibility of Cherenkov data with IVOA standards.

CTA is thus a perfect test case for the development of the IVOA Provenance data model. The CTA and IVOA communities are gathered together with other astronomy projects in the context of the ASTERICS project to develop common solutions.

\section{The IVOA Provenance data model and use cases for CTA}

\articlefigure{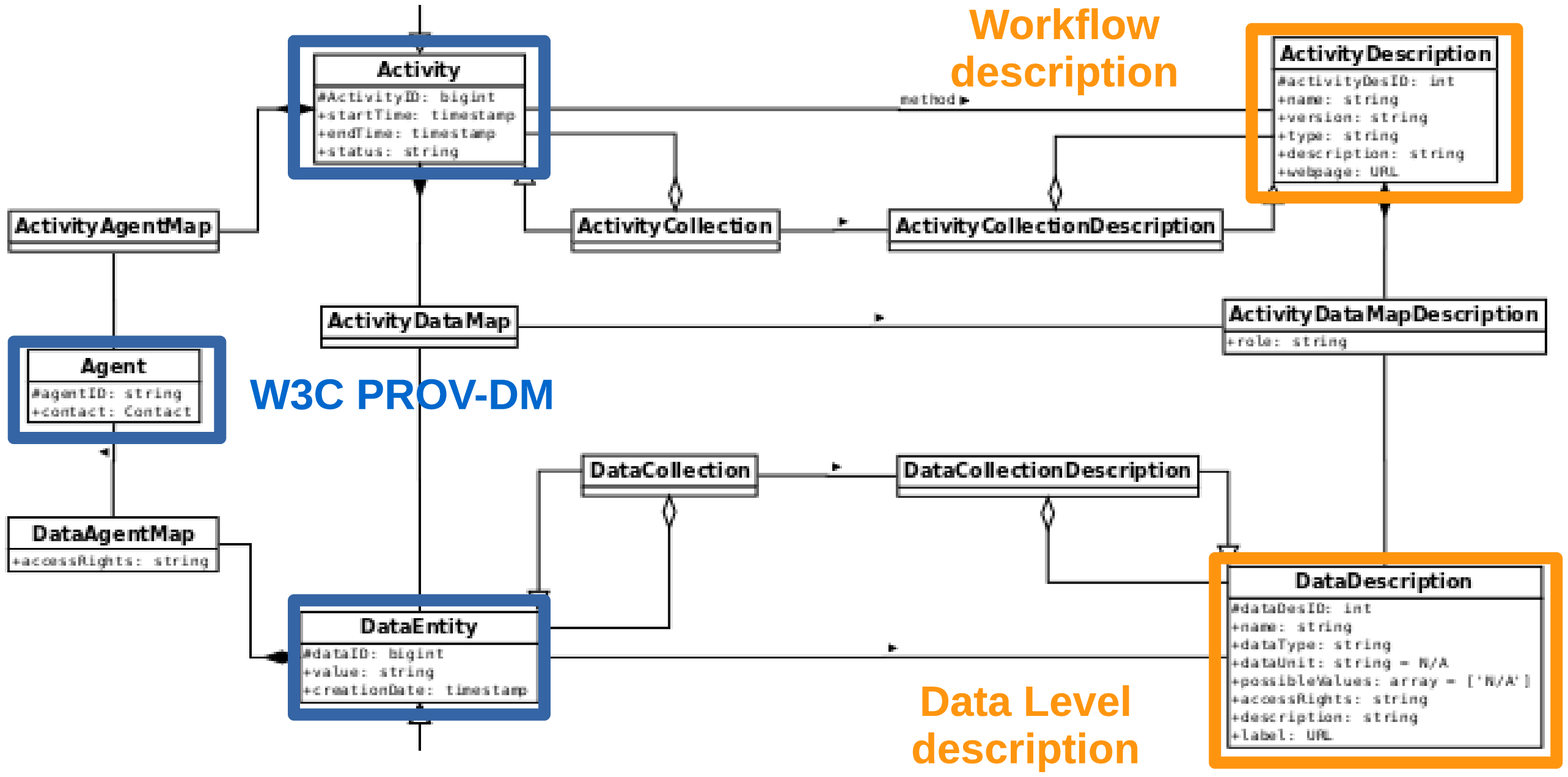}{fig3}{IVOA Provenance data model based on the W3C PROV data model, with the addition of a description side on the right that contains the available data levels and the pipeline workflow.}

The starting point of the IVOA Provenance data model is the PROV concept\footnote{\url{http://www.w3.org/TR/prov-overview/}} developed by the World Wide Web Consortium (W3C), based on three blocks and their relations: Entity, Activity (action on an Entity), and Agent (performing the Activity).

This general concept can be adapted to Astronomy projects where Entities are generally data products obtained from observations or simulations. Activities can be classified in different categories: observation preparation, data acquisition, data processing, data diffusion, publication... This concept was initially adapted to the RAVE (radial velocity experiment) survey by the German Astrophysical VO \citep{Riebe2015}.

We mainly focus here on the modeling of data processing Activities. Data acquisition Activities are generally very specific to each Astronomy project, making them harder to standardize.
We opt for a data model that includes a description side, as presented in Figure~\ref{fig3}, where data products and pipeline stages can be previously described. In this diagram, Activities and DataEntities can also be grouped as Collections. For each relation linking Activity-Entity-Agent, we added DataMaps to better describe many-many relations.
The key W3C concept is preserved in the IVOA Provenance, thus ensuring interoperability with available tools and services.

For CTA, one of the main requirement is thus to enable user queries on Provenance information, if possible in a standardized way. We identify several categories of use cases that will help to specify the IVOA Provenance implementation: 
1. Fill Provenance information during the processing (e.g. calling a function before and after each Pipeline stage).
2. Run quality checks and locate problems in the Pipeline.
3. Associate Provenance information to the data products.
4. Filter data products using Provenance criteria.
5. Reprocess a data product with different input parameters.

Those use cases will help to implement the different aspects of Provenance in Astronomy (e.g. data model, notation, access).

\section{Provenance notation and access}

We tested in particular the PROV-N notation to describe the different workflow stages. This can be easily converted to follow the PROV-XML schema which could be mapped to a VOTable or written as separate XML files. The use of PROV standards enables the compatibility with available services such as the generation of diagrams.

The access to Provenance information could be implemented using IVOA data access standards, such as the Table Access Protocol (TAP) and DataLink. We foresee that each data product type will have specific Provenance fields that could be queried. Those fields should probably be listed for each data product type in a Provenance profile. A Provenance service would thus first retrieve a Provenance profile and then create queries using this profile.

%Compatibility with UWS services?

\acknowledgements
ASTERICS (\url{http://www.asterics2020.eu/}) is a project supported by the European Commission Framework Programme Horizon 2020 Research and Innovation action under grant agreement n. 653477; Additional fundings by: CNRS/INSU (Action Spécifique Observatoire Virtuel, ASOV), Observatoire de Paris (Action Fédératrice CTA), and Paris Astronomical Data Centre.

%\bibliography{editor}  % For BibTex

% For non-BibTex:

\end{document}